\documentclass[aps,prd,twocolumn,superscriptaddress,showpacs,showkeys]{revtex4-1}
\usepackage{color}
\usepackage{fancyhdr}
\usepackage{amsmath}
\usepackage{amsfonts}
\usepackage{graphicx}
\usepackage{subfigure}
\usepackage{epstopdf}
\numberwithin{equation}{section}

\newcommand{\beq}[1]{\begin{equation}\label{#1}}
 \newcommand{\eeq}{\end{equation}}
 \newcommand{\bea}[1]{\begin{eqnarray}\label{#1}}
 \newcommand{\eea}{\end{eqnarray}}

\begin{document}
\title{Preinflationary Perturbations in The Thermal Background}
\author{Zucheng Li$^{1}$, Yanbin Deng$^{2}$, and Yong-chang Huang }
\affiliation{ Institute of Theoretical Physics, Beijing University of Technology, Beijing 100124, China\\
$^{2}$Department of Physics, Capital Normal University, Beijing, 100048,China}
\email{zucheng.li@emails.bjut.edu.cn, scientifichina@outlook.com, ychuang@bjut.edu.cn}

\date{\today}

\begin{abstract}
The solution to large-scale power deficit in the cosmic microwave background fluctuations may be relevant with the physics in the preinflationry period, and significantly depends on whether the initial conditions are set as Bunch-Davies vacuum or thermal ground state. We show the primordial cosmological perturbations based on the preinflationary model with three phases at different thermal initial conditions. The influences of different thermal initial states are studied by the detailed discussions of the adjusting role of model parameters on the power spectrum. Finally, this paper gives the solution to large-scale power deficit in the cosmic microwave background fluctuations.\\
\end{abstract}
\maketitle

\section{Introduction}

For the early universe, the standard cosmological model explains the cosmic microwave background (CMB) radiation and the formations of large-scale structures, but is plagued with some problems, such as horizon, flatness, entropy, homogeneity, isotropy and monopole\cite{1, 2, 3, 4}. For these problems, the slow-roll inflation model with single scaler field provided a successful solution. It gives the power spectrum for the density perturbations and gravitational waves originated from quantum fluctuations during the standard inflationary stage that successfully explains various observations on CMB radiation and the distributions of the galaxies in various large-scale structure observations\cite{5, 5.1, 5.2, 5.3, 6}. But recently updated observational data about the information of early universe from WMAP \cite{7, 8, 9} and Planck \cite{9.1, 9.2, 9.3, 9.4} point out that the standard slow-roll inflation can't explain the large-scale power deficit in CMB TT-mode spectrum, an increasingly indisputable result fortified by the continuously improving data reliability. Many models about preinflation are attempted at as possible explanation for this discrepancy. At least for the case when the slow-roll inflation last for just the minimal necessary e-folding numbers \cite{10}, the large-scale power spectrum anomalies may be attributed to the preinflationary non-slow-roll evolution.

Thus far, the continuous evolutions of the physical processes, for example the continuous evolutions of universe scale factor and the state parameters of fields, during the preinflationary stage of cosmic evolution, have not been successfully constructed. What researchers have been able to build are the models which divide the preinflationary period into multiple sections or phases for which the two-phase cases have been studied in detail\cite{32}. The problem about these models is the discontinuity of the field state parameters between adjacent phases, and the difficulty of continuously connecting the universe scale factor across adjacent phases. Our work analyzes in detail the models with three preinflationary phases to push forward the endeavor of more powerful model constructions. The freedom in the arrangement of the behavior of the state equations for the three-phase preinflationary models rewards the knowledge to distinguish the preinflationary phases into expanding and/or contracting stages. Thus, according to the expanding or contracting characters of the preinflationary phases, the three-phase preinflationary model we study can be divided into two types, the superinflation case and the bounce case. In the superinflation case, the universe is always expanding throughout the whole preinflationary stage, till it connects to the slow-roll inflation of the standard cosmological studies\cite{21, 21.1, 21.2, 21.3, 21.4, 21.5, 21.6, 21.7, 21.8}. In the bounce case, the universe is initially in a contracting phase, before it bounces into the expanding phase that again connects to the slow-roll inflation of the standard cosmological studies \cite{14, 15, 15.1, 16, 16.1, 16.2, 16.3, 16.4, 17, 17.1}. Both the two types of the models can explain the lager-scale power spectrum anomalies in CMB TT-mode spectrum by adjusting the model parameters.

A drawback in the usual calculations of the primordial perturbations in the multiple-phase preinflationary models is that, the initial condition for the quantization of the perturbations at the earliest preinflationary phase is always set as the Bunch-Davies vacuum (BD vacuum)\cite{11}. However, the evolution and the physics of the universe in the preinflationary period would unavoidably deflect the initial condition away from the naive considerations employed in the multiple-phase preinflationary models\cite{12, 13, 19}. With the improvements of the CMB temperature and polarization fluctuation observations, the fact that the primordial perturbations originated from different choices of the initial conditions might imprint significant influence on the CMB fluctuations at the large scales becomes a critical problem of consideration. The finiteness of the scale, energy density and temperature of the Planck stage, to which the earliest preinflationary phase is connecting as the onset for the cosmic expanding, forces us to set the thermal ground state as initial condition for perturbation generating in the superinflation case of the multiple- phase model. Similar arguments apply to the bounce cases of the multiple-phase models, because the scale factor of the universe in the initial contracting phase might contract to the previously mentioned Planck scale before it bounces into the expanding connecting to the slow-roll inflationary stage. When different ground states are adopted, it gives different results for perturbation power spectrum. When the Planck temperature and the Planck length are set as the initial temperature and the initial scale, the evolution of physical temperature of the inflaton field in each period of the multiple phase model and the physical temperature at the end of slow-roll inflation can be obtained\cite{30, 30.1, 30.2, 30.3, 30.4, 30.5, 30.6, 30.7}.

This paper is organized as follows. In Sect.\uppercase\expandafter{\romannumeral2}, we introduce the scale factor evolution equation and the equation of power spectrum. In Sect.\uppercase\expandafter{\romannumeral3}, we use power spectrum equation to compute the three kinds of three-phase models, and discuss the adjusting role of model parameters on the power spectrum. In Sect.\uppercase\expandafter{\romannumeral4}, we discuss the multiple phases model with thermal initial state, and compare the influence of different thermal initial states. In Sect.\uppercase\expandafter{\romannumeral5}, we discuss the temperature evolution of perturbation modes in three-phase models. In sect.\uppercase\expandafter{\romannumeral6}, there are the conclusion and discussion.

\section{Primordial perturbation of multiple phases model in BD vacuum}

Until now, the preinflationary physics are unclear. Researchers' various models explain our universe, but a lot of important problems still are remaining   open. As the stage before slow-roll inflation is very important, the preinflationary stage stimulats researchers to propose various reasonable models, which had been able to explained experiment data partly successfully. In this paper, we assume that the preinflationary stage is consisted of multiple phases with single field. We define the inflationary stage as phase 0, the nearest preinflationary phase as phase 1, so forth and so on. If there are multiple phases in preinflationary stage, we should consider the continuity of the adjacent phases.
\subsection{Multiple pases model of preinflation in BD vacuum}
As Ref. \cite{21}, the cosmological scale factor of phase i is
\beq{}
  a_{i} \sim \eta^{\frac{2}{1+3\omega_{i}}}, \qquad or \qquad (-\eta)^{\frac{2}{1+3\omega_{i}}},
\eeq
where $\eta=\int dt/a$ is the conformal time; $\omega_{i}=p_{i}/\rho_{i}$ ($\omega_{i}\neq -1/3$) is state parameter of phase i, with constant value. The scale factor evolutions of the inflationary stage and the phase i are given by Eq. (2.2) and Eq. (2.3) below. Because scale factor $a$ and Hubble constant $H$ are continuous transiting between adjacent phases, we require the following continuity conditions Eq. (2.4) and Eq. (2.5) between adjacent phases
\beq{}
 a_{0}(\eta) = \frac{a_{0}(\eta_{0})}{1 - \mathcal{H}_{inf}\cdot (\eta - \eta_{0})},
\eeq
\beq{}
 a_{i}(\eta) = a_{i}(\eta_{i-1})\left[ 1 +   \frac{1+3\omega_{i}}{2}\mathcal{H}_{i}(\eta_{i-1})\cdot(\eta-\eta_{i-1}) \right]^{\frac{2}{1+3\omega_{i}}},
\eeq
\beq{}
 a_{i}(\eta_{i-1})=a_{i-1}(\eta_{i-1}),
 \eeq
 \beq{}
 a_{i}^{'}(\eta_{i-1})=a_{i-1}^{'}(\eta_{i-1}),
\eeq
where, $\eta_{i}$ is the onset time of phase $i$, also the ending time of phase $i+1$. The prime denotes the derivative with respect to conformal time, and $\mathcal{H}_{inf}=\mathcal{H}_{0}(\eta_{0})$ is the conformal Hubble parameter of phase $0$. If we require that primordial perturbations are generated and exit the horizon in the phase $i$£¬it means the $|\mathcal{H}_{i}|$ had to increase with time. Thus the following conditions must be satisfied\cite{21, 21.1},
\begin{eqnarray}
 \omega_{i} & < -\frac{1}{3}  \qquad  for\quad the\quad expanding\quad phase,  \nonumber\\
 \omega_{i} &> -\frac{1}{3}  \qquad  for\quad the\quad contracting\quad phase.
\end{eqnarray}
Eq. (2.6) is the criterion for the situation of phase $i$, including slow-roll inflation. Without violating the null energy condition $\rho + p \geq 0$, we have $ -1 \leq \omega_{i} \leq 1$ for field potential energy $ V\geq 0 $; $\omega_{i} \leq -1 $ or $\omega_{i} \geq 1 $ for $ V\leq 0 $. However, if we violated the null energy condition, the value of $\omega_{i}$ can be any value expected for $\omega_{i}=1$. The violation of the null energy condition will lead to ghosts, which means that the the phases are instable, although it can possibly be avoided by higher order derivative scalar field, or modified gravity. We know that the bounce happens when $\omega$ goes from $\omega>-1/3$ to $\omega<-1/3$, which is instantaneous process between adjacent phases in multiple phases model. In the bounce cases, we can use the ekpyrotic model and cycle model to avoid ghost. In the superinflation, Ref. \cite{21.8} shows that there is no ghost instability. The state parameter of slow-roll inflation is $\omega \simeq -1$, and $\omega \ll -1$ is very slowly expanding.

\subsection{The power spectrum of preinflationary perturbations in BD vacuum}
As the curvature perturbation is generated in preinflationary stage with BD vacuum ground state, we should consider the effect of preinflationary stage. For the adjacent phases, we require it is continuous, and the variations between the adjacent phases are instantaneous, and the perturbations are continuous in preinflation expected to leave the horizon.

The equation of curvature perturbation derived from scalar field is \cite{41.493, 458.219}
\beq{}
 v^{''} + (c_{s}^{2}k^{2} - \frac{z^{''}}{z})v = 0   \quad,
\eeq
where $v \equiv z\mathcal{R}$, the prime is the derivative with respect to conformal time, $z \equiv a\sqrt{2M_{p}^{2}\epsilon } /c_{s} $ and $\epsilon \equiv -\dot{H}/H^{2}=3(1+\omega)/2$. In the superinflationary case, the absolute value of $\epsilon$ should be taken. $\mathcal{R}$ is perturbation curvature and we set $c_{s}^{2}=1$ for simplicity.

In slow-roll inflation, we have
\beq{}
\frac{z_{0}^{''}}{z_{0}} \simeq \frac{2\mathcal{H}_{inf}^{2}}{[1 - \mathcal{H}_{inf}(\eta - \eta_{0})]^{2}}.
\eeq
The solution of perturbation Eq. (2.8) is
\begin{small}
\beq{}
 v_{0}\! =\! \sqrt{-k\eta_{eff_{0}}}[C_{0,1}H^{(1)}_{3/2}(-k\eta_{eff_{0}}) \!+\! C_{0,2}H^{(2)}_{3/2}(-k\eta_{eff_{0}})] ,
\eeq
\end{small}
where ${\eta}_{eff_{0}} = \eta-\eta_{0}-\frac{1}{\mathcal{H}_{inf}}$, $H^{(1)}_{3/2}$ and $H^{(2)}_{3/2}$ is the $(3/2)$-th order Hankel function of the first and second kinds, respectively.

In the phase $i$ of preinflation, we have
\beq{}
\frac{z^{''}_{i}}{z_{i}} \simeq \frac{(1-3\omega_{i})\mathcal{H}^{2}_{i}(\eta_{i-1})}{2[1+\frac{1+3\omega_{i}}{2}\mathcal{H}_{i}(\eta_{i-1})\cdot (\eta - \eta_{i-1})]^{2}}.
\eeq
The solution of Eq. (2.10) is
\begin{small}
\beq{}
 v_{i}\! =\! \sqrt{-k\eta_{eff_{i}}}[C_{i,1}H^{(1)}_{\nu_{i}}(-k\eta_{eff_{i}})+C_{i,2}H^{(2)}_{\nu_{i}}(-k\eta_{eff_{i}})],   \eeq
 \end{small}
where $ \nu_{i} = \frac{3}{2}|\frac{1-\omega_{i}}{1+3\omega_{i}}| $, $\eta_{effi} = \eta - \eta_{i-1} + \frac{2}{(1+3\omega_{i})\mathcal{H}_{i}(\eta_{i-1})} $. $H^{(1)}_{\nu_{i}}$ and $H^{(2)}_{\nu_{i}}$ are the $\nu_{i}$-th order Hankel functions of the first kind and second kind, respectively.

When the perturbation mode $k$ goes through the adjacent phases, it must be continuous. Thus we have the recursive relationship for coefficients $C_{i,1}$ and $C_{i,2}$\cite{32}
\beq{}
\begin{aligned}
 \begin{pmatrix}
C_{i,1}\\
C_{i,2}
\end{pmatrix} &= \mathcal{M}^{i,i+1} \times \begin{pmatrix}
C_{i+1,1}\\
C_{i+1,2}
\end{pmatrix}\\
&= \mathcal{M}^{i,i+1} \times \mathcal{M}^{i+1,i+2} \times \dots \\
&\times \mathcal{M}^{i_{max}-1,i_{max}} \times \begin{pmatrix}
C_{i_{max},1}\\
C_{i_{max},2}
\end{pmatrix}  ,
\end{aligned}
\eeq
where $\mathcal{M}^{i,i+1}$ is the $2\times2$ matrix
\beq{}
\mathcal{M}^{i,i+1} = \begin{pmatrix}
\mathcal{M}_{11}  & \mathcal{M}_{12}  \\
\mathcal{M}_{21}  &  \mathcal{M}_{22}
\end{pmatrix},
\eeq
in which matrix element of $\mathcal{M}$ satisfy the relationship,
\begin{equation*}
\begin{aligned}
 \mathcal{M}_{11} &=\frac{i\pi\sqrt{xy}}{8}\{[-H_{-1+\nu_{i+1}}^{(1)}(y)+H_{1+\nu_{i+1}}^{(1)}(y)]H_{\nu_{i}}^{(2)}(x)\\
&+[H_{-1+\nu_{i}}^{(2)}(x)-H_{1+\nu_{i}}^{(2)}(x)]H_{\nu_{i+1}}^{(1)}(y)\\
&+(x^{-1}-y^{-1})H_{\nu_{i}}^{(2)}(x)H_{\nu_{i+1}}^{(1)}(y)\}  ,
\end{aligned}
\end{equation*}

\begin{equation*}
\begin{aligned}
\mathcal{M}_{12} &=\frac{i\pi\sqrt{xy}}{8}\{[-H_{-1+\nu_{i+1}}^{(2)}(y)+H_{1+\nu_{i+1}}^{(2)}(y)]H_{\nu_{i}}^{(2)}(x)\\
&+[H_{-1+\nu_{i}}^{(2)}(x)-H_{1+\nu_{i}}^{(2)}(x)]H_{\nu_{i+1}}^{(2)}(y)\\
&+(x^{-1}-y^{-1})H_{\nu_{i}}^{(2)}(x)H_{\nu_{i+1}}^{(2)}(y)\} ,
\end{aligned}
\end{equation*}

\begin{equation*}
\begin{aligned}
\mathcal{M}_{21}&=\frac{i\pi\sqrt{xy}}{8}\{[H_{-1+\nu_{i+1}}^{(1)}(y)-H_{1+\nu_{i+1}}^{(1)}(y)]H_{\nu_{i}}^{(1)}(x)\\
&-[H_{-1+\nu_{i}}^{(1)}(x)-H_{1+\nu_{i}}^{(1)}(x)]H_{\nu_{i+1}}^{(1)}(y)\\
&-(x^{-1}-y^{-1})H_{\nu_{i}}^{(1)}(x)H_{\nu_{i+1}}^{(1)}(y)\}  ,
\end{aligned}
\end{equation*}

\beq{}
\begin{aligned}
 \mathcal{M}_{22}&=\frac{i\pi\sqrt{xy}}{8}\{[H_{-1+\nu_{i+1}}^{(2)}(y)-H_{1+\nu_{i+1}}^{(2)}(y)]H_{\nu_{i}}^{(1)}(x)\\
&-[H_{-1+\nu_{i}}^{(1)}(x)-H_{1+\nu_{i}}^{(1)}(x)]H_{\nu_{i+1}}^{(2)}(y)\\
&-(x^{-1}-y^{-1})H_{\nu_{i}}^{(1)}(x)H_{\nu_{i+1}}^{(2)}(y)\} ,
\end{aligned}
\eeq
where $x=-\frac{2k}{(1+3\omega_{i})\mathcal{H}_{i}(\eta_{i})}$,$ y=-\frac{2k}{(1+3\omega_{i+1})\mathcal{H}_{i+1}(\eta_{i})}$.

In the phase $i_{max}$, according to Eq. (2.10), $|\eta_{i_{max}}|$ is very large, we have the approximate condition $k^{2} \gg \frac{z^{''}_{imax}}{z_{imax}}$, then the solution of Eq. (2.7) is
\beq{}
  u \sim \frac{1}{\sqrt{2k}}e^{-ik\eta} ,
\eeq
where the coefficients of phase $i_{max}$ are
\beq{}
   C_{imax,1}=\frac{\sqrt{\pi}}{2\sqrt{k}} ,\qquad    C_{imax,2}=0 .
\eeq
Coefficients $C_{i,1}$ and $C_{i,2}$ satisfy the canonical normalization constraint
\beq{}
   \frac{4k}{\pi}(|C_{i,1}|^{2} - |C_{i,2}|^{2})=1 .
\eeq
The corresponding power spectrum of curvature perturbations mode $k$ is
\beq{}
  \mathcal{P}_{\mathcal{R}} = \frac{k^{3}}{2\pi^{2}}|\frac{v_{0}}{z_{0}}|^{2} .
\eeq
Substituting Eq. (2.9) into Eq. (2.18), and requiring $|k(\eta-\eta_{0})-\frac{k}{\mathcal{H}_{inf}}| \ll 1 $, we get a universal formula
\beq{}
 \mathcal{P}_{\mathcal{R}} =  \mathcal{P}_{\mathcal{R}}^{inf} \frac{4}{\pi}k|C_{0,1}-C_{0,2}|^{2},
\eeq
where $\mathcal{P}_{\mathcal{R}}^{inf}=\frac{1}{2M_{p}^{2}}(\frac{H_{inf}}{2\pi})^{2}$ is the standard slow-roll inflation spectrum, and $H_{inf}=\frac{\mathcal{H}_{0}(\eta_{0})}{a_{0}(\eta_{0})}$ is the Hubble constant of inflation, $\mathcal{H}_{inf}=\mathcal{H}_{0}(\eta_{0})$. The spectral index of the scalar perturbations in multiple phases model is,
\beq{}\begin{aligned}
 n_{s} &=1+\frac{dln\mathcal{P}_{\mathcal{R}}}{dlnk} = 1+\frac{dln\mathcal{P}_{\mathcal{R}}^{inf}}{dlnk}+\frac{dln(k|C_{0,1}-C_{0,2}|^{2})}{dlnk}\\
 &= n_{inf}+\frac{dln(k|C_{0,1}-C_{0,2}|^{2})}{dlnk} ,
\end{aligned}\eeq
where, $n_{inf}$ is the spectral index of the slow-roll inflation, and $n_{inf}$ is nearly unity in the data of WMAP.

\section{Three phase model in BD vacuum}
In this section, we use the universal formula Eq. (2.19) to get the power spectrum of the multiple phase model. Actually, if we only discuss the three phase model, according to Eq. (2.3), the three phase model can be divided into two kinds, the case of superinflationary and the cases with bounce. In superinflationary case, the state parameters of all three-phase satisfy $\omega_{i}<-1$, e.g. , every phase is expanding. Ref. \cite{22}, the reasonability of superinflationary case had been discussed clearly. Matter with the parameter $\omega<-1$ is named as phantom matter. The expansion with $\omega \ll -1$ is slow, while for $\omega \sim -1$ it is rapid and can be regarded as phantom inflation \cite{35, 35.1, 35.2, 35.3, 35.4}. The reasonability of the cases with bounce had been discussed clearly in Ref. \cite{23}. For the ekpyrotic or cyclic model without violating the null energy condition, $\omega \gg 1$, and then we can set $\omega \gg 1$ in our cases with ghost-free \cite{25}. The contracting with $\omega \gg 1$ is rapid. In the first type of bounce, only the third phase is contracting, and the rest of phases are expanding. In the second type of bounce, the third phase and the second phase are contracting, the first phase is expanding. It is worth noting that all the cases, we discussed, are singularity-free \cite{31, 31.1, 31.2, 31.3, 31.4, 31.5}. By computing different cases, we get different power spectra.

In three phase model, $i_{max}=3$, we get the solutions of Eq. (2.9) and Eq. (2.11)
\begin{equation*}
\begin{aligned}
v_{3} &=\sqrt{-k\eta+k\eta_{2}-\frac{2k}{(1+3\omega_{3})\mathcal{H}_{3}(\eta_{2})}} \cdot \frac{\sqrt{\pi}}{2\sqrt{k}} \\
 &\cdot H^{(1)}_{\nu_{3}}(-k\eta+k\eta_{2}-\frac{2k}{(1+3\omega_{3})\mathcal{H}(\eta_{2})})  \\
&=\frac{\sqrt{\pi}}{2}\sqrt{-\eta+\eta_{2}-\frac{2}{(1+3\omega_{3})\mathcal{H}_{3}(\eta_{2})}} \cdot \\
& H^{(1)}_{\nu_{3}}(-k\eta+k\eta_{2}-\frac{2k}{(1+3\omega_{3})\mathcal{H}_{3}(\eta_{2})})
\end{aligned}
\end{equation*}
\begin{equation*}
\begin{aligned}
v_{2} &=\sqrt{-k\eta+k\eta_{1}-\frac{2k}{(1+3\omega_{2})\mathcal{H}_{2}(\eta_{1})}} \\
& [C_{2,1}H^{(1)}_{\nu_{2}}(-k\eta+k\eta_{1}-\frac{2k}{(1+3\omega_{2})\mathcal{H}_{2}(\eta_{1})})+ \\
& C_{2,2}H^{(2)}_{\nu_{2}}(-k\eta+k\eta_{1}-\frac{2k}{(1+3\omega_{2})\mathcal{H}_{2}(\eta_{1})}) ] ,
\end{aligned}
\end{equation*}
\begin{equation*}
\begin{aligned}
v_{1} &=\sqrt{-k\eta+k\eta_{0}-\frac{2k}{(1+3\omega_{1})\mathcal{H}_{1}(\eta_{0})}}  \\
&[C_{1,1}H^{(1)}_{\nu_{1}}(-k\eta+k\eta_{0}-\frac{2k}{(1+3\omega_{1})\mathcal{H}_{1}(\eta_{0})})+ \\
&C_{1,2}H^{(2)}_{\nu_{1}}(-k\eta+k\eta_{0}-\frac{2k}{(1+3\omega_{1})\mathcal{H}_{1}(\eta_{0})})] ,
\end{aligned}
\end{equation*}
\beq{}
\begin{aligned}
v_{0}&=\sqrt{-k\eta+\frac{k}{\mathcal{H}_{inf}} }[  C_{0,1}H^{(1)}_{3/2}(-k\eta+\frac{k}{\mathcal{H}_{inf}})\\
&+ C_{0,2}H^{(2)}_{3/2}(-k\eta+\frac{k}{\mathcal{H}_{inf}})],
\end{aligned}
\eeq
where, the conformal Hubble parameters are
\begin{eqnarray}
 &\mathcal{H}_{1}(\eta)=\mathcal{H}_{inf}\left[ 1+\frac{1+3\omega_{1}}{2}\mathcal{H}_{inf}\cdot \eta \right]^{-1} ,\nonumber  \\
 &\mathcal{H}_{2}(\eta)=\mathcal{H}_{2}(\eta_{1})\left[ 1+\frac{1+3\omega_{2}}{2}\mathcal{H}_{2}(\eta_{1})\cdot (\eta-\eta_{1}) \right]^{-1}\nonumber , \\
 &\mathcal{H}_{3}(\eta)=\mathcal{H}_{3}(\eta_{2})\left[ 1+\frac{1+3\omega_{3}}{2}\mathcal{H}_{3}(\eta_{2})\cdot (\eta-\eta_{2}) \right]^{-1}   .
\end{eqnarray}
By computing the recursive matrix, we get the solutions of every phase, and then compute the specific power spectrum expression of every case. To compare the spectrum of this case with the spectrum of slow-roll inflation directly, we define
\beq{}
  P(\eta_{2},\eta_{1},\omega_{3},\omega_{2},\omega_{1})=\frac{\mathcal{P}_{\mathcal{R}}}{\mathcal{P}_{\mathcal{R}}^{inf}}
   = \frac{4}{\pi}k|C_{0,1}-C_{0,2}|^{2}  ,
\eeq
where $P(\eta_{2},\eta_{1},\omega_{3},\omega_{2},\omega_{1})$ is the shape of the power spectrum, $C_{0,1}$ and $C_{0,2}$ are computed from $C_{3,1}$ and $C_{3,2}$ by the following recursive matrix
\beq{}
\begin{aligned}
 \begin{pmatrix}
C_{0,1}\\
C_{0,2} \end{pmatrix} = M^{0,1}\times M^{1,2}\times M^{2,3} \times \begin{pmatrix}
C_{3,1}\\
C_{3,2} \end{pmatrix} \\
  = M^{0,1}\times M^{1,2}\times M^{2,3} \times \begin{pmatrix}
\frac{\sqrt{\pi}}{2\sqrt{k}}\\
0 \end{pmatrix} .
\end{aligned}
\eeq
Next, we will compute the power spectra of specific cases, and plot $P(\eta_{2},\eta_{1},\omega_{3},\omega_{2},\omega_{1})$ with specific cases in three phase cases. So we can observe the influence of parameters in three phase model on spectra, and the $\mathcal{H}_{inf}$ is determined by the amplitude of CMB fluctuation.
\subsection {The superinflatinary case in BD vacuum}
For an expanding universe, we know $\omega_{1}, \omega_{2}, \omega_{3}<-\frac{1}{3}$, and all the $\omega_{i}<-1$ define the superinflationary case.
Due to the continuity of conformal Hubble parameter in superinflationary case, we have
\begin{small}
\beq{}
\mathcal{H}_{1}(\eta_{0})=\mathcal{H}_{0}(\eta_{0}), \mathcal{H}_{2}(\eta_{1})=\mathcal{H}_{1}(\eta_{1}), \mathcal{H}_{3}(\eta_{2})=\mathcal{H}_{2}(\eta_{2}).
\eeq
\end{small}

\begin{figure*}[htbp]
\centering
\subfigure[the effect of $\eta$]
{
\begin{minipage}[t]{0.48\linewidth}
\centering
\includegraphics[scale=0.48]{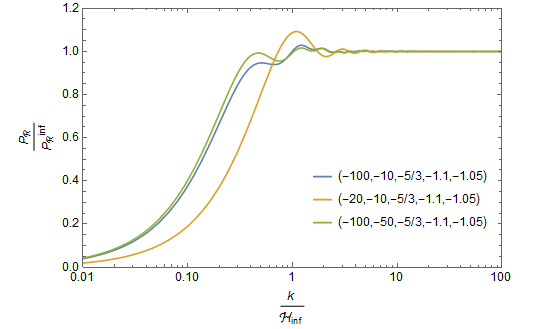}
\end{minipage}%
}%
\subfigure[the effect of $\omega$]
{
\begin{minipage}[t]{0.48\linewidth}
\centering
\includegraphics[scale=0.46]{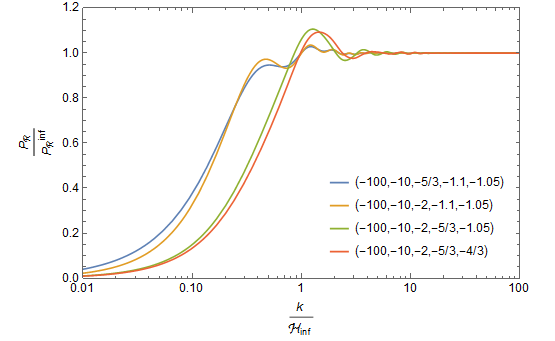}
\end{minipage}%
}%
\centering
\caption{the effect of parameters on spectrum in superinflation. Those parameters in figure.1 is ($\eta_{2}$,$\eta_{1}$,$\omega_{3}$,$\omega_{2}$,$\omega_{1}$) }
\end{figure*}
According to Eq. (3.3), Eq. (3.4) and Eq. (3.5), we can get the specific power spectrum corresponding to different values of the parameters ($\eta_{2}$, $\eta_{1}$, $\omega_{3}$, $\omega_{2}$, $\omega_{1}$). In figure 1, we compare the influences of parameters in three-phase model. In figure 1(a),
with the increased $|\eta_{2}|$ value, the blue curve is shown to have a lower valued first peak than that of the orange curve, and also a left shift. With the increased $|\eta_{1}|$ value, the green curve is shown to have a higher valued first peak than that of the blue curve. In figure 1(b), with the increased $|\omega_{3}|$ value, the blue curve is shown to have a lower valued first peak than that of the orange curve, and also a left shift. With the increased $|\omega_{2}|$ value, the green curve is shown to have a higher valued first peak than that of the orange curve, and also a right shift. With the increased $|\omega_{1}|$ value, the red curve is shown to have a slightly right shift than that of the green curve. In the three-phase model, the effect of parameters is complicated. In figure 1(a), we only change one parameter at different curves. When we change the $|\eta_{2}|$, it means the phase 2 period will be longer, because $|\eta_{2}|$ is very big. But when change the $|\eta_{1}|$, it means periods of phase 1 and phase 2 will be change at the same curves. Thus we will choose the perfect rank for parameters by comparing with the CMB data.

\subsection{The first type bounce case in BD vacuum}
In the first type bounce case, $\omega_{3}>-1/3$, and $\omega_{2}, \omega_{1}, \omega_{0}<-1/3$. Due to the continuity in the first bounce case, we have
\begin{small}
\beq{}
\mathcal{H}_{1}(\eta_{0})=\mathcal{H}_{0}(\eta_{0}),\mathcal{H}_{2}(\eta_{1})=\mathcal{H}_{1}(\eta_{1}),   \mathcal{H}_{3}(\eta_{2})=-\mathcal{H}_{2}(\eta_{2}).
\eeq
\end{small}
\begin{figure*}[htbp]
\centering
\subfigure[the effect of $\eta$]{
\begin{minipage}[t]{0.47\linewidth}
\centering
\includegraphics[scale=0.52]{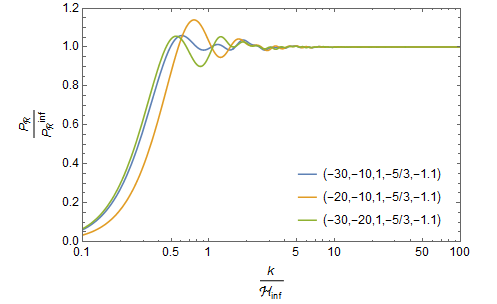}
\end{minipage}%
}%
\subfigure[the effect of $\omega$]{
\begin{minipage}[t]{0.49\linewidth}
\centering
\includegraphics[scale=0.51]{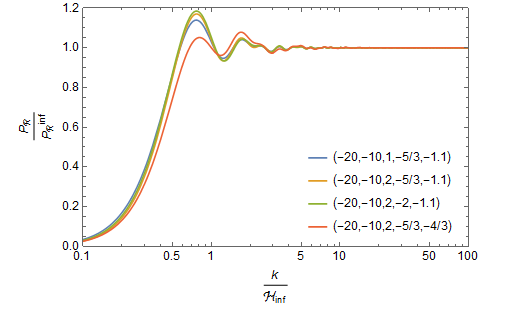}
\end{minipage}%
}%
\centering
\caption{the effect of parameters on spectrum in the first type bounce case. Those parameters in figure.2 is ($\eta_{2}$,$\eta_{1}$,$\omega_{3}$,$\omega_{2}$,$\omega_{1}$)}
\end{figure*}

In the following figure 2(a), with the increased $|\eta_{2}|$ value, the blue curve is shown to have a lower valued first peak than that of the orange curve, and also a left shift. With the increased $|\eta_{1}|$ value, the green curve is shown to have a left shift first peak than that of the blue curve. In figure 2(b), with the increased $|\omega_{3}|$ value, the blue curve is shown to have a lower valued first peak than that of the orange curve. With the increased $|\omega_{2}|$ value, the green curve is shown to have a higher valued first peak than that of the orange curve. With the increased $|\omega_{1}|$ value, the red curve is shown to have a lower valued first peak than that of the orange curve. In the three-phase model, the effect of parameters is complicated. Thus we will choose the perfect rank for parameters by comparing with the CMB data.

\subsection{The second type bounce case in BD vacuum}
In the second type bounce case, we know $\omega_{3}, \omega_{2}>-1/3$, and $\omega_{1}, \omega_{0}<-1/3$. Due to the continuity, we have
\begin{small}
\beq{}
\mathcal{H}_{1}(\eta_{0})=\mathcal{H}_{0}(\eta_{0}), \mathcal{H}_{2}(\eta_{1})=-\mathcal{H}_{1}(\eta_{1}), \mathcal{H}_{3}(\eta_{2})=\mathcal{H}_{2}(\eta_{2}).
\eeq
\end{small}
\begin{figure*}[htbp]
\centering
\subfigure[the effect of $\eta_{i}$]{
\begin{minipage}[t]{0.47\linewidth}
\centering
\includegraphics[scale=0.52]{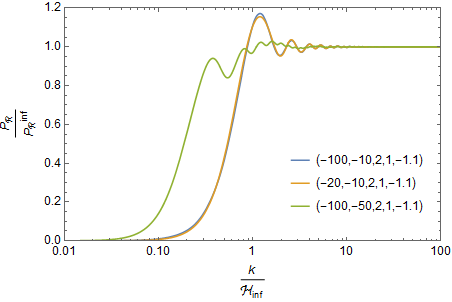}
\end{minipage}%
}%
\subfigure[the effect of $\omega_{i}$]{
\begin{minipage}[t]{0.46\linewidth}
\centering
\includegraphics[scale=0.52]{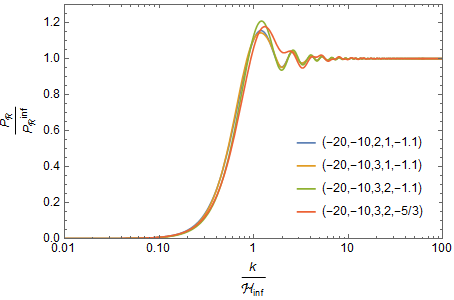}
\end{minipage}%
}%
\centering
\caption{the effect of parameters on spectrum in the second type bounce case. Those parameters in figure.3 is ($\eta_{2}$,$\eta_{1}$,$\omega_{3}$,$\omega_{2}$,$\omega_{1}$)}
\end{figure*}
In figure 3(a), with the increased $|\eta_{2}|$ value, the blue curve is shown to have a higher valued first peak than that of the orange curve. With the increased $|\eta_{1}|$ value, the green curve is shown to have a lower valued first peak than that of the blue curve, and also a left shift. In figure 3(b), with the increased $|\omega_{3}|$ value, the blue curve is shown to have a lower valued first peak than that of the orange curve. With the increased $|\omega_{2}|$ value, the green curve is shown to have a higher valued first peak than that of the orange curve, and also a right shift. With the increased $|\omega_{1}|$ value, the red curve is shown to have a lower valued first peak than that of the blue curve, and also a right shift. In the three-phase model, the effect of parameters is complicated. Thus we will choose the perfect rank for parameters by comparing with the CMB data.

Generally, depending on the value of $\omega$, the three phase model consists of superinflation, the first type of bounce and the second type of bounce. We attempt to study the influence of the multiple phase model on preinflation, but different with Ref. \cite{32}. We compute the three phases, then we find that the influence of every parameter is not simple as the two phase model. In two phase model, one parameter only can affect the first peak or the shift of spectrum, but in the three phase model, one parameter can affect the first peak and the shift of spectrum at the same time. The multiple phase model is significance for the large-scale power deficit in CMB, but there are some problems that the variations between the adjacent phases are instantaneous and continuous, and the influences of the parameters are complicated which one can not describe clearly. However, there are so many types of bounce and superinflation in the model, which have more than three phases, that the calculation will be hard to do. We want to get an equation which can describe the variations of $\omega$ and $\eta$, but it needs more calculations for bounce case and superinflationary case, so that we can not do the multiple phase model well. In addition, we can join the thermal background into the multiple phase model, which can get a relationship between the multiple phase model and the thermal temperature.\\

\section{The multiple phase model with thermal initial state }
In the above sections, we introduce the multiple phase model in  BD-vacuum to solve the large-scale power deficit, and compute the perturbations and power spectra of three phase model with specific parameters. In the superinflation case of the multiple phase model, as the onset for the cosmic expanding, the phase $i_{max}$ follows from the Planck stage, which is a physical state with finite scale energy density and temperature, thus we should consider the thermal ground state for generating perturbation. In the bounce cases of the multiple phase model, the scale factor $a$ is contracting at first and then expanding to the slow-roll inflationary stage. That the ekpyrotic model, as one of the solutions to avoid the ghost appearing in some of the bounce scenario, has a corresponding ekpyrotic temperature \cite{27}, points to us that we should also set the thermal ground state for bounce cases. In this section, we compute the perturbation power spectrum on thermal ground state. The quantum statistics tells us, the effects of annihilation operator and creation operator acting on thermal ground state are different from acting on vacuum state. With thermal state, we should consider the effect of boson particle number. In the computing process, we find that the model with the thermal ground state contribute only a multiplying factor to the BD-vacuum power spectrum, common for both of the bounce and superinflationary cases. Next, we quantize the operator $\hat{v}$, and discuss the influence of thermal ground state. \\

\subsection{The perturbation quantization under thermal initial state }
Quantizing the operator $\hat{v}_{k}$, we have
\beq{}
 v_{k} \rightarrow \hat{v}_{k}(\eta)=v_{k}(\eta)\hat{a}_{k} + v^{\ast}_{-k}(\eta)\hat{a}^{\dagger}_{-k},
\eeq
where the annihilation operator and creation operator of mode k, $(\hat{a}_{k},\hat{a}_{-k}^{\dag})$, satisfy the commutation relation
\beq{}
 [\hat{a}_{k},\hat{a}^{\dagger}_{k^{'}}]=(2\pi)^{3}\delta^{3}(k-k^{'}).
 \eeq
Operators $\hat{v}_{k}$ satisfy the relation
\beq{}
\frac{i}{\hbar}(v^{\ast}_{k}v^{'}_{k}-v^{\ast'}_{k}v_{k})=1 .
\eeq
In the universal evolution, for every perturbation mode k, the annihilation operator and creation operator of the initial state have relevance with the annihilation operator and creation operator of the final state
\beq{}
\hat{b}_{k}=c_{+}(k)\hat{a}_{k}+c_{-}^{\ast}(k)\hat{a}_{-k}^{\dag},\quad\hat{b}_{k}^{\dag}=c_{-}(k)\hat{a}_{-k}+c_{+}^{\ast}(k)\hat{a}_{k}^{\dag},
\eeq
where $(\hat{a}_{k},\hat{a}_{-k}^{\dag})$ are the annihilation operator and creation operator of initial state $|in>$, and $(\hat{b}_{k},\hat{b}_{-k}^{\dag})$
are the annihilation operator and creation operator of final state $|out>$. In Eq. (4.4), the values of Bogoliubov coefficients $c_{\pm}$ depend on cosmological background geometrical dynamics, and satisfy $|c_{+}|^{2}-|c_{-}|^{2}=1$. In our paper, we set the beginning of our model in state $|in>$, and the ending in state $|out>$. It's worth noting that we get a simple factor which shows the influence of thermal initial state on power spectrum, by comparing BD vacuum case and thermal initial state case.

When the state $|in>$ is BD vacuum state, we have $\hat{a}_{k}|in>=\hat{a}_{k}|0>=0$. The corresponding particle number of state $|out>$ is
\beq{}
 \bar{N}_{k}=<0|\hat{b}^{\dagger}_{k}\hat{b}_{k}|0>=|c_{k}|^{2} .
\eeq
The power spectrum of scalar field perturbation is
\beq{}
 <0|\hat{\mathcal{R}}_{k}\hat{\mathcal{R}}_{k^{'}}|0> = \frac{v_{k}^{2}}{z^{2}} .
\eeq
Actually we compute the spectrum of $v_{0}$.
\beq{}
 \mathcal{P}_{\mathcal{R}}=\frac{k^{3}}{2\pi^{2}}|\mathcal{R}_{k}|^{2}
=\frac{k^{3}}{2\pi^{2}}|\frac{v_{k}^{2}}{z^{2}}|^{2} .
\eeq
Eq. (4.7) can be written as
\beq{}
 \mathcal{P}_{\mathcal{R}}=\frac{k^{3}}{2\pi^{2}}|\frac{v_{0}}{z_{0}}|^{2}  .
\eeq
When the initial state is  thermal initial state, but not BD vacuum, we have
\beq{}
\hat{a}^{\dagger}_{k}\hat{a}_{k}|in>= \hat{a}^{\dagger}_{k}\hat{a}_{k}|\mathcal{T}>=n_{k}|\mathcal{T}>  ,
\eeq
where particle number $n_{k}=\frac{1}{e^{k/\mathcal{T}}-1}$, $\mathcal{T}$ is conformal temperature of initial state. Thus the particle number of state $|out>$ has been changed \cite{24}
\beq{}
\begin{aligned}
 \bar{N}_{k}&=<\mathcal{T}|\hat{b}^{\dagger}_{k}\hat{b}_{k}|\mathcal{T}>=|c_{-}(k)|^{2}(1+n_{k}) + n_{k}(1+|c_{-}(k)|^{2})\\
  &\simeq |c_{-}(k)|^{2}coth\left(\frac{k}{2\mathcal{T}}\right) .
  \end{aligned}
\eeq
The particle number of thermal initial state get a factor $coth\left(\frac{k}{2\mathcal{T}}\right) $ more than BD vacuum initial state. We also easily get the perturbation spectrum of scalar field in thermal initial state
\beq{}
  <\mathcal{T}|\hat{\mathcal{R}}_{k}\hat{\mathcal{R}}_{k^{'}}|\mathcal{T}> =|\frac{v_{0}}{z_{0}}|^{2}coth\left(\frac{k}{2\mathcal{T}}\right),
\eeq
\beq{}
 \mathcal{P}_{\mathcal{R}}=\frac{k^{3}}{2\pi^{2}}|\frac{v_{0}}{z_{0}}|^{2}coth\left(\frac{k}{2\mathcal{T}}\right)  .
 \eeq
Obviously, the perturbation spectrum of thermal initial state has a factor $coth(\frac{k}{2\mathcal{T}})$ more than BD vacuum state. The above quantized process, which is independent with the preinflationary model, is valid for boson on thermal ground state. Thus we can use the same process to compute the power spectrum of the multiple phase model on thermal ground state and the specific three-phase cases on thermal ground state with conformal temperature $\mathcal{T}$.

\subsection{The power spectrum of three-phase model with thermal initial state }
In the above discussion, we get the power spectrum of multiple phases on thermal ground state by multiplying the power spectrum of BD vacuum state with $coth(\frac{k}{2\mathcal{T}})$. Thus we discuss the effect of the thermal ground state on the three phase model with specific parameter values
\beq{}
\begin{aligned}
 &P(\eta_{2},\eta_{1},\omega_{3},\omega_{2},\omega_{1},\mathcal{T})=\frac{\mathcal{P}_{\mathcal{R}}}{\mathcal{P}_{\mathcal{R}}^{inf}}\\
   &= \frac{4}{\pi}k|C_{0,1}-C_{0,2}|^{2}\coth\left(\frac{k}{2\mathcal{T}}\right).
   \end{aligned}
\eeq
The factor $coth\left(\frac{k}{2\mathcal{T}}\right)$ at right-hand side, coming from thermal ground state, enhances the power spectrum at large-scale. We can constrain the value of $\mathcal{T}$ by comparing the power spectrum with CMB observations.

\begin{figure*}[htbp]
\centering
\subfigure[superinflation]{
\begin{minipage}[t]{0.33\linewidth}
\centering
\includegraphics[scale=0.4]{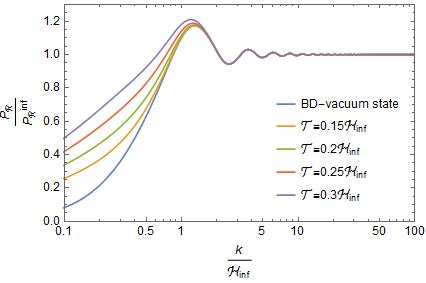}
\end{minipage}%
}%
\subfigure[the first type bounce ]{
\begin{minipage}[t]{0.33\linewidth}
\centering
\includegraphics[scale=0.37]{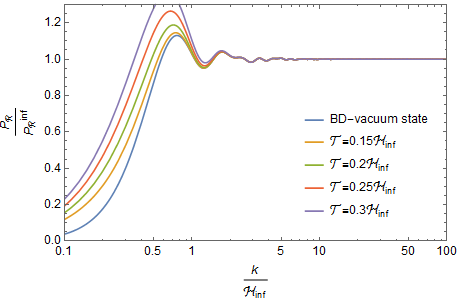}
\end{minipage}%
}%
\subfigure[the second type bounce ]{
\begin{minipage}[t]{0.33\linewidth}
\centering
\includegraphics[scale=0.38]{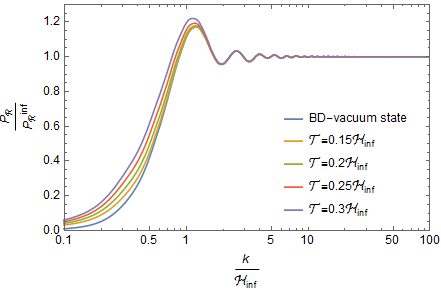}
\end{minipage}%
}%
\centering
\caption{comparing the three phases model in BD-vacuum state with thermal state, the parameters of the multiple phases model is ($\eta_{2}$, $\eta_{1}$, $\omega_{3}$, $\omega_{2}$, $\omega_{1}$). In (a), superinflation with $(-100, -10, -3, -7/3, -1.05)$; in (b), the first type bounce with $(-20, -10, 1, -5/3, -1.1)$; in (c), the second type bounce with $(-100, -10, 2, 1, -1.1, -1)$.  }
\end{figure*}

In Figure. 4, we compare the BD-vacuum ground states with different thermal ground state for the three phase cases. We want to get an appropriate temperature range for every case. In the figure. 4, for the specific parameters, when $\mathcal{T}<0.15\mathcal{H}_{inf}$, compared with the BD-vacuum ground state, the effect of thermal ground state on the first peak of power spectrum can be ignored. When the conformal temperature is higher than $0.15\mathcal{H}_{inf}$, the power spectrum will be enhanced, and with the increase of the conformal temperature $\mathcal{T}$, the enhancement of the power spectrum is more significant. This is common for all the three cases i.e., superinflationary case, the first type of bounce case and the second type of bounce case. We use thermal ground state and the three-phase model of preinflation to explain the large-scale power deficit in CMB observations. Therefore, for the three phase model with specific parameters, we have a corresponding range of conformal temperature $\mathcal{T}$, which will make the power spectrum agree well with CMB observations. It is important to adjust the parameters of the three phase model at a given specific conformal temperature $\mathcal{T}$ of the thermal ground state, for the power spectrum of the thermal ground state to agree with the observation.

\section{The temperature evolution of inflaton perturbations }

Now, we discuss the physical temperature evolution $T(\eta)$ of the multiple phase model in the thermal ground state. In discussing the scalar field, we have $\mathcal{T}=aT$ for the massless inflaton, and the $\mathcal{T}$ is constant. Thus, in the multiple phases model of preinflation, we have
\beq{}
\mathcal{T}=a(\eta)T(\eta)=a_{i}(\eta)T_{i}(\eta)=constant,
\eeq
where, we know $T\sim \frac{1}{a}$, from Eq. (2.6) so we have
\begin{eqnarray}
 \omega_{i} &< -\frac{1}{3},  \quad T_{i}\quad is\quad descending\quad in\quad  phase\quad  i   \nonumber ,\\
\omega_{i} &> -\frac{1}{3},    \quad T_{i}\quad is\quad increasing\quad in\quad  phase\quad  i,
\end{eqnarray}
where $T_{i}$ is the physical temperature of inflatons in phase $i$. Because the value of $\mathcal{T}$ is unclear, the value of $T_{i}$ is unclear. If we know the values of physical temperature and scale factor at a given time, we know the value of conformal temperature. In multiple phase model, considering Eq. (5.1) for the preinflationary stage and inflationary stage, we have
\beq{}
 \frac{T_{0}(\eta_{end})}{T_{i}(\eta)}=\frac{a_{i}(\eta)}{a_{0}(\eta_{end})}=\frac{a_{i}(\eta)}{a_{0}(\eta_{0})}\times \frac{a_{0}(\eta_{0})}{a_{0}(\eta_{end})}  ,
\eeq
where, $a_{i}(\eta)$ is the evolving scale factor in preinflationary phase i, the $\eta_{0}$ is the onset of inflation, $\eta_{end}$ is the ending of inflation and the onset of reheating, and $T_{0}(\eta_{end})$ is the physical temperature at the ending of inflation. In Eq. (5.3), we get the relation of $T_{i}(\eta)$ in preinflationary stage with the $\frac{a_{i}(\eta)}{a_{0}(\eta_{0})}$, because
\beq{}
\frac{a_{0}(\eta_{0})}{a_{0}(\eta_{end})} \simeq  e^{-N} ,
\eeq
\beq{}
 \eta_{end} \simeq \frac{1-e^{-N}}{\mathcal{H}_{inf}} .
\eeq
We get the relation between $\eta_{end}$ and $\mathcal{H}_{inf}$, if we have the specific value of e-folding number $N$, which is constrained to be above a lower bound determined by the CMB observations. After substituting Eq. (5.4) into Eq. (5.3) , we have
\beq{}
 \frac{a_{i}(\eta)}{a_{0}(\eta_{end})} \simeq \frac{a_{i}(\eta)}{a_{0}(\eta_{0})} \times e^{-N}.
\eeq
Substituting Eq. (5.6) into Eq. (5.3), we have
\beq{}
T_{i}(\eta) \simeq T_{0}(\eta_{end}) \times \frac{a_{0}(\eta_{0})}{a_{i}(\eta)} \times e^{N} .
\eeq
From Eq. (2.2) and Eq. (2.3), we have
\beq{}
\frac{a_{i}(\eta)}{a_{0}(\eta_{0})}=\frac{a_{i}(\eta)}{a_{i-1}(\eta_{i-1})} \times \frac{a_{i-1}(\eta_{i-1})}{a_{i-2}(\eta_{i-2})}  \times \cdots \times \frac{a_{1}(\eta_{1})}{a_{0}(\eta_{0})} ,
\eeq
where, for a general term $\frac{a_{j}(\eta)}{a_{j-1}(\eta_{j-1})}$, we have
\beq{}
\frac{a_{j}(\eta)}{a_{j-1}(\eta_{j-1})} = \left[ 1+\frac{1+3\omega_{j}}{2}\mathcal{H}_{j}(\eta_{j-1}) \cdot (\eta-\eta_{j-1})  \right]^{\frac{2}{1+3\omega_{j}}} ,
\eeq
and substituting Eq. (5.9) into Eq. (5.8), we have
\beq{}
\begin{aligned}
\frac{a_{i}(\eta)}{a_{0}(\eta_{0})}&= \left[ 1+\frac{1+3\omega_{i}}{2}\mathcal{H}_{i}(\eta_{i-1}) \cdot (\eta-\eta_{i-1})  \right]^{\frac{2}{1+3\omega_{i}}} \\
&\times \left[ 1+\frac{1+3\omega_{i-1}}{2}\mathcal{H}_{i-1}(\eta_{i-2}) \cdot (\eta_{i-1}-\eta_{i-2})  \right]^{\frac{2}{1+3\omega_{i-1}}}\\
&\times \cdots \times \left[ 1+\frac{1+3\omega_{1}}{2}\mathcal{H}_{1}(\eta_{0}) \cdot (\eta_{1}-\eta_{0})  \right]^{\frac{2}{1+3\omega_{1}}}  ,
\end{aligned}
\eeq

where, the Eq. (5.10) is applicable for cases of any number phases. The relations between the conformal Hubble parameters $\mathcal{H}_{i}(\eta)$ in Eq. (5.10) are given as
\begin{eqnarray}
 &\mathcal{H}_{0}(\eta)= \frac{\mathcal{H}_{inf}}{1-\mathcal{H}_{inf}\cdot\eta},  \nonumber\\
&\mathcal{H}_{i}(\eta) = \frac{\mathcal{H}_{i}(\eta_{i-1})}{1+\frac{1+3\omega_{i}}{2}\mathcal{H}_{i}(\eta_{i-1})\cdot (\eta-\eta_{i-1})}.
\end{eqnarray}
From here, we can get $\mathcal{H}_{i}(\eta_{i})$ from $\mathcal{H}_{o}(\eta_{0})$. Thus, it can be seen that $T_{i}(\eta)$ is dependent on the parameters $\eta_{i}$, $\omega_{i}$ and the values of $\mathcal{H}_{inf}$ and $N$.

Now we discuss the specific three-phase model with thermal initial state. For $i=3$,
\beq{}
T_{3}(\eta) \simeq T_{0}(\eta_{end}) \times \frac{a_{0}(\eta_{0})}{a_{3}(\eta)} \times e^{N} .
\eeq
In the superinflation cases, the onset of phase 3 is the ending of Planck stage, so the value of $T_{3}(\eta_{3})$ is about the scale of Planck temperature $T_{p}=m_{p}c^{2}/k_{b}$ with $m_{p}=\sqrt{\frac{\hbar c}{G}}$, and the value of $a_{3}(\eta_{3})$ is about the scale of Planck length $l_{p}=\sqrt{\frac{\hbar G}{c^{3}}}$. Thus the value of conformal temperature is $\mathcal{T} = a_{3}(\eta_{3}) T_{3}(\eta_{3})  \sim \hbar c/k_{b}$. For the specific case with given values of parameters, with a given value $N$, we get the approximate value of the physical temperature $T_{o}(\eta_{end})$, which is the temperature of inflatons at the onset of reheating. From Eq. (5.12), one has
\beq{}
 T_{0}(\eta_{end}) \simeq \frac{a_{3}(\eta_{3})T_{3}(\eta_{3})}{a_{0}(\eta_{0})} \cdot e^{-N} \simeq  \frac{\hbar c}{k_{b} a_{0}(\eta_{end})} .
\eeq
In the bounce cases, if the scale factor $a(\eta)$ shrinks to Planck scale before the onset of the earliest expanding phase, by the same mathematics as the superinflation cases, we get the same physical temperature $T_{o}(\eta_{end})\simeq  \frac{\hbar c}{k_{b} a_{0}(\eta_{end})}$. If the universe does not shrink to Planck scale before the onset of the earliest expanding phase, according to the Eq. (5.12), the value of $T_{o}(\eta_{end})$ will depend on the physical temperature and the scale factor of the universe at the onset of the earliest expanding phase of specific model.

\section{Conclusion and Discussion}
The large-scale power deficit in CMB TT-mode spectrum implies that there may exist some unclear periods before inflation. The multiple phase model\cite{32}, discussed above, can suppress the power spectrum on large-scale mode. In our work, according to the general framework of the multiple phase model, we analyse the three phase model, which introduces three phase in the preinflationary period. For the multiple phase model of single field, the evolution equation of scale factor $a_{i}$ for phase $i$ is constraint by the state parameter $\omega_{i}$. Thus, depending on whether the phase $i$ is expanding or contracting, the multiple phases model can be divided into two types, the superinflationary case and the bounce case. In the three phase model of preinflation, we have one superinflationary case and two bounce cases. By comparing the shapes of the ratio function between the power spectrum of three phase model and the power spectrum of slow-roll inflation, we get the relation of the large-scale power deficit in CMB TT-mode spectrum with the parameters of the specific three phase model. In the three phase model, being generated in phase 3, and leaving the horizon when $k=\mathcal{H}_{inf}$, the perturbation mode is continuous between adjacent phases. In the equation of curvature perturbation, we set $c_{s}^{2}=1$ for simplicity. Then we compare the power spectrum at different parameter values in BD vacuum to get the appropriate value for CMB TT-mode spectrum.

As the onset for the cosmic expansion in the superinflationary case of the multiple phase model, the phase $i_{max}$ follows from the Planck stage. Because the Planck stage is a physical state with finite scale energy density and temperature, it requires us to consider the thermal ground state for perturbation generating. The scale factor $a$ is contracting at first before expanding to the slow-roll inflationary stage in the bounce cases. For the bounce cases, we also consider the thermal ground state. When different ground states are being acted upon by the creation and annihilation operator of each mode k, it gives different results for perturbation power spectrum. For the perturbation quantization in the thermal ground state, we work with the ground state corresponding to non-zero particle number of bosons. Then we find that the introduction of the thermal ground state with finite conformal temperature contributes only a multiplying factor to the BD-vacuum power spectrum, common for both of the bounce and superinflationary cases. The introduction of thermal ground state will enhance the power spectrum for the large-scale mode, but it does not influence the power spectrum for the small-scale mode. For the three phase model with specific parameters, there is a corresponding range of conformal temperature $\mathcal{T}$ that makes the power spectrum agree well with CMB observations.

According to the conformal temperature at the perturbation production, we can get the evolution of physical temperature $T_{i}$ of inflaton in each period of the multiple phase model. In the superinflationary case of three phase model, the Planck temperature and the Planck length being set as the initial temperature and the initial scale, the physical temperature at the end of slow-roll inflation can be obtained by the scale factor evolution\cite{30}. In the bounce cases, the physical temperature at the end of slow-roll inflation will depend on the physical temperature and the scale factor of the universe at the onset of the earliest expanding phase of specific model\cite{31}.

Some relevant problems are worthy of further considerations. For the perfect fluid, $c_{s}^{2}=\omega_{i}$, in the phase $i$, where $\omega_{i}$ is constant. The equation of curvature perturbation with $c_{s}^{2}$ has been studied\cite{28, 29}, and we expect to study the effect of $c_{s}^{2}$ on adjacent phases and the influence on the power spectrum in preinflationary period. We can study the influence of the scalar field energy on the reheating process in the framework of multiple phase model. Also we can study the influence of curvature perturbation on cosmic anisotropies after slow-roll inflation in multiple phase model with BD vacuum state and the thermal initial states. It is hoped to find an equation for describing the continuous variation of state parameters in preinflationary stage that hence can be applied to describe more natural variation of the scale factor along the whole process of the universe evolution. The model describing the contraction of the universe needs more detailed discussions. The prospects of accommodating multiple-source observational data requirements for the framework of multiple-phase inflation model present very promising.


\begin{thebibliography}{99}
\bibitem{1}A. H. Guth, {\em Phys. Rev. D} {\bf23}, 347 (1981).
\bibitem{2}A. A. Starobinsky, {\em Phys. Lett.} {\bf91B}, 99 (1980).
\bibitem{3}A. D. Linde, {\em Phys. Lett.} {\bf108B}, 389 (1982).
\bibitem{4}A. Albrecht and P. J. Steinhardt, {\em Phys. Rev. Lett.} {\bf48}, 1220 (1982).
\bibitem{5}L. P. Grishchuk, {\em Annals N. Y. Acad. sci.} {\bf302} (1977) 439.
\bibitem{5.1}L. P. Grishchuk, {\em Sov. Phys. JETP} {\bf40} (1975) 409.
\bibitem{5.2}L. P. Grishchuk, {\em Annals JETP Lett.} {\bf23}(1976) 293 .
\bibitem{5.3}A. A. Starobinsky, {\em JETP Lett.} {\bf30} (1979) 682.
\bibitem{6}D. H. Lyth and A. Riotto, {\em Physics Reports} {\bf314}, 1 (1999).
\bibitem{7}D. N. Spergel et al. (WMAP Collaboration), {\em Astrophysical. J. Suppl.} {\bf148}, 175 (2003).
\bibitem{8}E. Komatsu et al. (WMAP Collaboration), {\em Astrophysical. J. Suppl.} {\bf192}, 18 (2011).
\bibitem{9}G. Hinshaw et al. (WMAP Collaboration), {\em Astrophysical. J. Suppl.} {\bf208}, 19 (2013).
\bibitem{9.1}P. A. R. Ade et al. (Planck Collaboration), {\em Astron. Astrophys.} {\bf571}, A1 (2014).
\bibitem{9.2}P. A. R. Ade et al. (Planck Collaboration), {\em Astron. Astrophys.} {\bf571}, A22 (2014).
\bibitem{9.3}P. A. R. Ade et al. (Planck Collaboration), {\em Astron. Astrophys.} {\bf571}, A16 (2014).
\bibitem{9.4}P. A. R. Ade et al. (Planck Collaboration), {\em Astron. Astrophys.} {\bf594}, A11 (2016).
\bibitem{10}E. Ramirez and D. J. Schwarz, {\em Phys. Rev. D} {\bf85}, 103516 (2012).
\bibitem{32}Y. Cai, Y. T. Wang, and Y. S. Piao, {\em Phys. Rev. D} {\bf92}, 023518 (2015).
\bibitem{21}Y. S. Piao and Y. Z. Zhang, {\em Phys. Rev. D} {\bf70}, 043516 (2004).
\bibitem{21.1}Y. S. Piao, {\em Phys. Lett. B} {\bf606}, 245 (2005).
\bibitem{21.2}M. Baldi, F. Finelli, and S. Matarrese, {\em Phys. Rev. D} {\bf72}, 083504 (2005).
\bibitem{21.3}Y. S. Piao and E. Zhou, {\em Phys. Rev. D} {\bf68}, 083515 (2003).
\bibitem{21.4}P. Creminelli, A. Nicolis, and E. Trincherini, {\em J. Cosmol. Astropart. Phys.} {\bf11} (2010) 021.
\bibitem{21.5}K. Hinterbichler, A. Joyce, J. Khoury, and G. E. J. Miller, {\em J. Cosmol. Astropart. Phys.} {\bf12} (2012) 030.
\bibitem{21.6}K. Hinterbichler, A. Joyce, J. Khoury, and G. E. J. Miller, {\em Phys. Rev. Lett.} {\bf110}, 241303 (2013).
\bibitem{21.7}Y.S. Piao, {\em Phys. Lett. B} {\bf701}, 526 (2011).
\bibitem{21.8}Z.-G. Liu, J. Zhang, and Y.-S. Piao, {\em Phys. Rev. D} {\bf84}, 063508 (2011).
\bibitem{41.493}V. F. Mukhanov, {\em JETP Lett.} {\bf41}, 493 (1985).
\bibitem{458.219}J. Garriga and V. F. Mukhanov, {\em Phys. Lett. B} {\bf458}, 219 (1999).
\bibitem{14}Y.-S. Piao, B. Feng, and X.-m. Zhang, {\em Phys. Rev. D} {\bf69}, 103520 (2004).
\bibitem{15}Y.-S. Piao, {\em Phys. Rev. D} {\bf71}, 087301 (2005).
\bibitem{15.1}Y.-S. Piao, S. Tsujikawa, and X.-m. Zhang, {\em Classical Quantum Gravity} {\bf21}, 4455 (2004).
\bibitem{16}B. A. Powell and W. H. Kinney, {\em Phys. Rev. D} {\bf76}, 063512 (2007).
\bibitem{16.1}F. T. Falciano, M. Lilley, and P. Peter, {\em Phys. Rev. D} {\bf77}, 083513 (2008).
\bibitem{16.2}M. Lilley, L. Lorenz, and S. Clesse, J, {\em Cosmol. Astropart. Phys.} {\bf06}, 004 (2011).
\bibitem{16.3}J. Mielczarek, {\em J. Cosmol. Astropart. Phys.} {\bf1} (2008) 011.
\bibitem{16.4}J. Mielczarek, M. Kamionka, A. Kurek, and M. Szydlowski, {\em J. Cosmol. Astropart. Phys.} {\bf07} (2010) 004.
\bibitem{17}Z. G. Liu, H. Li, and Y. S. Piao, {\em Phys. Rev. D} {\bf90}, 083521 (2014).
\bibitem{17.1}T. Biswas and A. Mazumdar, {\em Classical Quantum Gravity} {\bf31}, 025019 (2014).
\bibitem{11}D. Baumann, {\em arXiv:0907.5424}.
\bibitem{12}C.R. Contaldi, M. Peloso, L. Kofman, and A.D. Linde, {\em J. Cosmol. Astropart. Phys. } {\bf07} (2003) 002.
\bibitem{13}J. M. Cline, P. Crotty, and J. Lesgourgues, {\em J. Cosmol. Astropart. Phys.} {\bf09} (2003) 010.
\bibitem{19}Y. T. Wang and Y. S. Piao, {\em Phys. Lett. B} {\bf741}, 55 (2015).
\bibitem{30}L.P. Grishchuk, {\em Space Sci. Rev.} {\bf148}, 315 (2009).
\bibitem{30.1}L.P. Grishchuk, {\em Class. Quant. Grav.} {\bf10}, 2449 (1993).
\bibitem{30.2}B.A. Powell and W.H. Kinney, {\em Phys. Rev. D} {\bf76}, 063512 (2007).
\bibitem{30.3}I.Chin Wang and K.-W. Ng, {\em Phys. Rev. D} {\bf77}, 083501 (2008).
\bibitem{30.4}G. Marozzi, M. Rinaldi, and R. Durrer, {\em Phys. Rev. D} {\bf83}, 105017 (2011).
\bibitem{30.5}K. Bhattacharya, S. Mohanty, and R. Rangarajan, {\em Phys. Rev. Lett.} {\bf96}, 121302 (2006).
\bibitem{30.6}S. Das, G. Goswami, J. Prasad, and R. Rangarajan, {\em J. Cosmol. Astropart. Phys.} {\bf06} (2015) 001.
\bibitem{30.7}S. Das, G. Goswami, J. Prasad, and R. Rangarajan, {\em Phys. Rev. D} {\bf93}, 023516 (2016).
\bibitem{22}Z. G. Liu, Z. K. Guo, and Y. S. Piao, {\em Eur. Phys. J. C} {\bf74}, 3006 (2014).
\bibitem{35}M. H. Namjoo, H. Firouzjahi, and M. Sasaki, {\em J. Cosmol. Astropart. Phys.} {\bf12} (2012) 018.
\bibitem{35.1}M. Cicoli, S. Downes, and B. Dutta, {\em J. Cosmol. Astropart. Phys.} {\bf12} (2013) 007.
\bibitem{35.2}F. G. Pedro and A. Westphal, {\em J. High Energy Phys.} {\bf04} (2014) 034.
\bibitem{35.3}R. K. Jain, P. Chingangbam, J. O. Gong, L. Sriramkumar, and T. Souradeep, {\em J. Cosmol. Astropart. Phys.} {\bf01} (2009) 009.
\bibitem{35.4}R. K. Jain, P. Chingangbam, L. Sriramkumar, and T.Souradeep, {\em Phys. Rev. D} {\bf82}, 023509 (2010).
\bibitem{23}Y. Cai, Y. T. Wang, J. Y. Zhao, and Y. S. Piao, {\em Phys. Rev. D} {\bf97}, 103535 (2018).
\bibitem{25}D. Battefeld and P. Peter, {\em Physics Reports} {\bf571}, 1 (2015).
\bibitem{31}J. Quintin, Y.F. Cai, and R.H. Brandenberger, {\em Phys. Rev. D} {\bf90}, 063507 (2014).
\bibitem{31.1}Y. F. Cai, T. Qiu, Y. S. Piao, M. Li, and X. Zhang, {\em J. High Energy Phys.} {\bf10} (2007) 071.
\bibitem{31.2}C. Lin, R. H. Brandenberger, and L. Perreault Levasseur, {\em J. Cosmol. Astropart. Phys.} {\bf04} (2011) 019.
\bibitem{31.3}Y. F. Cai, D. A. Easson, and R. Brandenberger, {\em J. Cosmol. Astropart. Phys.} {\bf08} (2012) 020.
\bibitem{31.4}Y. F. Cai, E. McDonough, F. Duplessis, and R. H. Brandenberger, {\em J. Cosmol. Astropart. Phys.} {\bf10} (2013) 024.
\bibitem{31.5}Y.-F. Cai, R. Brandenberger, and P. Peter, {\em Classical Quantum Gravity} {\bf30}, 075019 (2013).
\bibitem{27}J. Khoury, B. A. Ovrut, P. J. Steinhardt, and N.Turok, {\em Phys. Rev. D} {\bf64}, 123522.
\bibitem{24}M. Gasperini, M. Giovannini, and G. Veneziano, {\em Phys. Rev. D} {\bf48}, R439 (1993).
\bibitem{28}M. Nakashima, R. Saito, Y. i. Takamizu, and J. Yokoyama, {\em Prog. Theor. Phys.} {\bf125}, 1035 (2011).
\bibitem{29}H. Firouzjahi and M. H. Namjoo, {\em Phys. Rev. D} {\bf90}, 063525 (2014).



\end{thebibliography}
\end{document}